\documentclass[reprint,twoside,superscriptaddress,aps,pra,longbibliography]{revtex4-2}
\usepackage[english]{babel}
\usepackage[utf8]{inputenc}
\usepackage[colorinlistoftodos, color=green!40, prependcaption]{todonotes}
\usepackage{siunitx}
\usepackage{svg}
\usepackage{graphicx}
\usepackage[caption=false]{subfig}
\usepackage[T1]{fontenc}
\newcommand\thefont{\expandafter\string\the\font}
\usepackage{amsmath}
\usepackage{mathtools}
\usepackage{mathrsfs}
\usepackage{mathptmx}
\usepackage{float}

\begin{document}
\title{Resonance frequency tracking schemes for micro- and nanomechanical resonators}

\author{Hajrudin Bešić}\email{hajrudin.besic@tuwien.ac.at}
\affiliation{Institute of Sensor and Actuator Systems, TU Wien, Gusshausstrasse 27-29, 1040 Vienna, Austria}
\author{Alper Demir}
\affiliation{Department of Electrical Engineering, Koç University, Istanbul 34450, Turkey}
\author{Johannes Steurer}
\affiliation{Institute of Sensor and Actuator Systems, TU Wien, Gusshausstrasse 27-29, 1040 Vienna, Austria}
\author{Niklas Luhmann}
\affiliation{Institute of Sensor and Actuator Systems, TU Wien, Gusshausstrasse 27-29, 1040 Vienna, Austria}
\author{Silvan Schmid}
\affiliation{Institute of Sensor and Actuator Systems, TU Wien, Gusshausstrasse 27-29, 1040 Vienna, Austria}

\date{\today} 

\begin{abstract}
Nanomechanical resonators can serve as high performance detectors and have potential to be widely used in the industry for a variety of applications. Most nanomechanical sensing applications rely on detecting changes of resonance frequency. In commonly used frequency tracking schemes, the resonator is driven at or close to its resonance frequency. Closed-loop systems can continually check whether the resonator is at resonance and accordingly adjust the frequency of the driving signal. In this work, we study three resonance frequency tracking schemes, a feedback-free (FF), a self-sustaining oscillator (SSO), and a phase-locked loop oscillator (PLLO) scheme. We improve and extend the theoretical models for the FF and the SSO tracking schemes, and test the models experimentally with a nanoelectromechanical system (NEMS) resonator. We employ a SSO architecture with a pulsed positive feedback topology and compare it to the commonly used PLLO and FF schemes. We show that all tracking schemes are theoretically equivalent and that they all are subject to the same speed versus accuracy trade-off characteristics. In order to verify the theoretical models, we present experimental steady-state measurements for all tracking schemes. Frequency stability is characterized by computing the Allan deviation. We obtain almost perfect correspondence between the theoretical models and the experimental measurements. These results show that the choice of the tracking scheme is dictated by cost, robustness and usability in practice as opposed to fundamental theoretical differences in performance.
\end{abstract}


\maketitle
 \message{The column width is: 8,6cm}
\section{Introduction} \label{sec:outline}
    Micro- and nanomechanical resonators are exceptional sensors whose resonance frequency detunes with parameter changes, which can be a change in mass \cite{chaste2012nanomechanical,hanay2012single,burg2007weighing,lee2010toward,schmid2013real}, change in damping \cite{khan2013online,bose2014micromechanical}, or change in stiffness \cite{larsen2011ultrasensitive,chien2018single,kurek2017nanomechanical}.    
    Resonance frequency tracking can be accomplished via open- and closed-loop schemes:
    \begin{itemize}  
    \item The feedback free (FF) or open-loop approach: In the standard configuration, this scheme has an inferior speed performance when compared to closed-loop and self-adjusting approaches, since it is limited by the mechanical response time of the resonator.    \item Phase-locked loop oscillator (PLLO) approach: It is commonly used since it can be easily realized digitally with DSP/FPGA based setups. Having a wide frequency range, one setup can be suitable for most micro- and nanomechanical resonators. There are many commercial devices available that can be used for PLLO implementations.
    \item Self-sustaining oscillator (SSO) with positive feedback approach: The classical realization of the SSO found limited use, mostly because they typically are implemented as analog circuits resulting in a narrow frequency range.
    \end{itemize}
    
    The frequency tracking schemes above have been compared both theoretically \cite{Demir} and experimentally \cite{Pedram}, where it was concluded that with the closed-loop PLLO and the self-adjusting SSO schemes better speed performance can be obtained at the expense of degraded precision in the presence of significant transduction noise, as compared to the FF approach. We show that one can in fact obtain the same speed versus precision trade-off using the open-loop FF scheme with a simple modification to the standard configuration. The theory behind these approaches have been discussed in depth in \cite{Demir}. We also extend the SSO model from \cite{Demir} by taking into account the impact of detection noise through two separate mechanisms and show that the resulting frequency fluctuations are equivalent to the PLLO including the detection noise limited regime. 
    
    There are various designs for the implementation of the positive feedback mechanism in the SSO configuration. The most common is to simply amplify and adjust the phase of the resonator response signal to generate the feedback drive, resulting in a sine-driven SSO. Alternatively, amplitude and duration adjustable timed pulses can be used in forming the drive for the resonator with a positive feedback mechanism, resulting in a pulse-driven SSO \cite{dominguez2005novel,schmid2008electrostatically,brenes2016influence,juillard2019towards,juillard2021experimental,gorreta2012pulsed}.  We designed and realized an SSO configuration based on the pulsed feedback mechanism. To experimentally investigate the SSO and the PLLO approaches, we compare our SSO implementation to a DSP/FPGA based lock-in amplifier setup with an integrated PLL system.

\section{Theory} \label{sec:develop}
\subsection{Noise in nanomechanical resonators}
We consider two noise sources, the \textit{thermomechanical} noise of the resonator and the \textit{detection} noise generated in the transduction and detection hardware. 

The thermomechanical noise is the most fundamental noise source of the resonator. It can be modeled as a white noise force $n_\mathrm{th}(t)$ at the input of the resonator \cite{silvanbook,Demir} and has a one sided spectral density with units [N$^2$/Hz]
\begin{equation}
    S_\mathrm{F}=\frac{4m\omega_0k_\mathrm{B}T}{Q},
\end{equation}
where $k_\mathrm{B}$ is Boltzmann's constant, $T$ the temperature, $m$ the effective mass,  $\omega_0$ the eigenfrequency, and $Q$ is the quality factor of the resonator. For a slightly damped resonator, the eigenfrequency is approximately equal to the resonance frequency $\omega_\mathrm{r}\approx\omega_0$. The white noise force at the input of the resonator is shaped by the resonator's complex force susceptibility
\begin{equation}
    \chi(s)= \frac{1/m}{s^2+ \frac{\omega_0}{Q} s+\omega^2_0}
\end{equation}
resulting in a steady-state thermomechanical amplitude noise at its output with a power spectral density
\begin{equation}
    S_\mathrm{th}(\omega)=S_\mathrm{F} \; \lvert\chi(\mathrm{j} \omega)\rvert^2.
\end{equation} 
Operating at resonance, the thermomechanical amplitude noise reduces to
\begin{equation}
    S_\mathrm{th}=S_\mathrm{th}(\omega_0)= S_\mathrm{F} \left(\frac{Q}{m \omega_0^2}\right)^2 = \frac{4 k_\mathrm{B}T Q}{m \omega_0^3}.
\end{equation}

Detection noise $n_d(t)$ is produced during the conversion of the mechanical motion into an electrical signal. It includes the noise generated by the readout and electronics. Detection noise can be modeled as a white noise source with respect to thermomechanical amplitude noise
\begin{equation}
    S_\mathrm{d} = \mathcal{K}^2 S_\mathrm{th},
\end{equation}
where $\mathcal{K}$ is a dimensionless factor that corresponds to the ratio between the detection noise background and the height of the thermomechanical noise peak.
If $\mathcal{K}>1$, the thermomechanical noise peak is buried in detection noise. If $\mathcal{K}<1$, thermomechanical noise is resolved above the detection noise background.
    
The resonator is driven by a coherent force $F(t)$, which results in a steady state amplitude response at resonance of $x(t) = F(t) Q/(m \omega_0^2) = A_\mathrm{r} \cos{(\omega_0 t)}$ with an amplitude $A_\mathrm{r}$.
Both, thermomechanical and detection amplitude noise then translates into corresponding phase noise with the following power spectral densities \cite{demirfundamental}
\begin{equation}\label{eq:phasenoise}
\begin{split}
    S_{\theta_\mathrm{th}} &= \frac{2}{A_\mathrm{r}^2} S_\mathrm{th}\\
    S_{\theta_\mathrm{d}} &= \frac{2}{A_\mathrm{r}^2} S_\mathrm{d} = \mathcal{K}^2 S_{\theta_\mathrm{th}}.
\end{split}
\end{equation}
This transformation clearly shows that a large oscillation amplitude $A_\mathrm{r}$ dilutes the effect of thermomechanical and detection noise. The maximization of $A_\mathrm{r}$ is limited by the onset of nonlinearities in the resonator system.

The conversion of the phase noise (\ref{eq:phasenoise}) into frequency noise $S_{\Delta \omega}(\omega)$ depends on the tracking scheme and will be derived in the following sections.
    

\subsection{Feedback-free (FF) scheme}

\begin{figure}
\centering
\includegraphics[clip,width=\columnwidth]{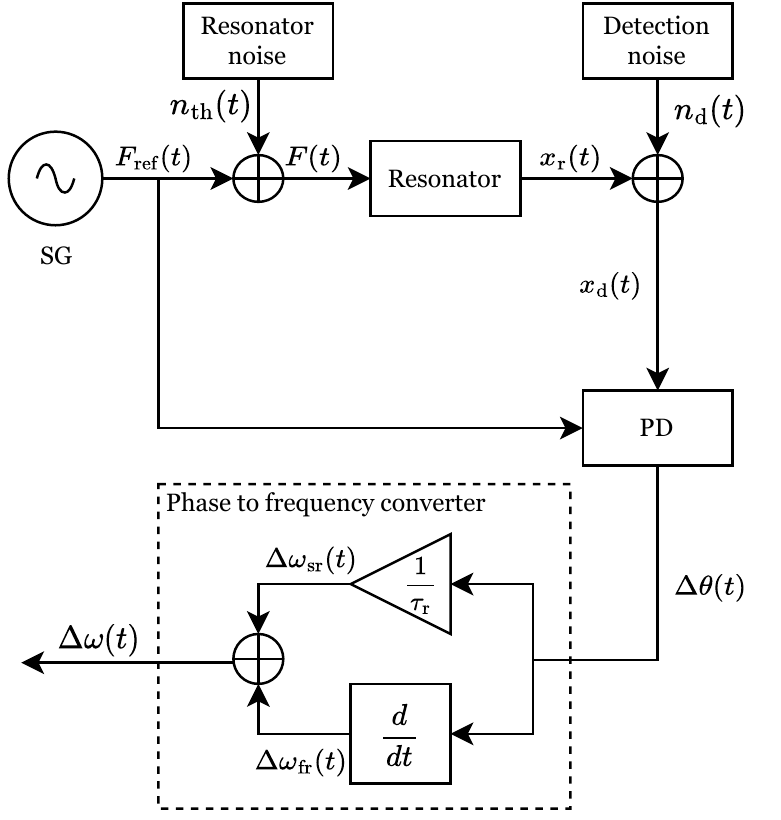}
\caption{Block diagram of the feedback-free frequency tracking scheme featuring a signal generator (SG) that is driving a micro- or nanomechanical resonator. The resonator response is fed into a phase difference detector (PD). The phase difference signal is then mapped to a frequency signal by the phase to frequency conversion mechanism. The thermomechanical noise is added at the input of the resonator and the detection noise is added at the input of the phase detector.}
\label{fig:ff_block}
\end{figure}
    
The feedback-free scheme is a simple and well known method where the resonator is driven at or close to resonance. Figure \ref{fig:ff_block} shows a schematic representation of the FF frequency tracking scheme. It consists of a signal generator that is driving the resonator with a constant frequency close to its resonance frequency. The resonator motion is then transduced and  the phase difference of the resonator response with respect to the reference signal provided by the driving signal generator is obtained by a phase detector (PD). Any sudden change in the resonance frequency, $\Delta \omega_\mathrm{r}$,  will cause a corresponding change in the detected phase difference $\Delta \theta(t)$, as derived in \cite{Demir} as follows 
\begin{equation}
\Delta \theta(t)=\tau_\mathrm{r} \Delta \omega_\mathrm{r} 
\left( 1-e^{-t/\tau_\mathrm{r}} \right),
\end{equation}
where
\begin{equation}\label{eq:responsetime}
    \tau_\mathrm{r}=\frac{2\,Q}{\omega_\mathrm{r}}
\end{equation}
is the mechanical resonator time constant. The resonance frequency change $\Delta \omega_\mathrm{r}$ can be easily extracted from this phase response. The drawback of the FF scheme is that the drive frequency has to be within the line-width of the resonator where the phase response is linear. This makes this scheme susceptible to thermal drift \cite{Demir, FM_demodulation}. Here, we assume that the drive signal indeed satisfies this condition for the FF scheme.
    
The standard map from the phase difference to the resonance frequency change is obtained simply by a division with $\tau_\mathrm{r}$, as described in \cite{Demir} and implemented in \cite{Pedram}. The frequency response in this case, which we call the {\it slow response}, is given by
\begin{equation}
\Delta \omega_\mathrm{sr}(t)= \frac{\Delta \theta(t)}{\tau_\mathrm{r}} =
\Delta \omega_\mathrm{r} 
\left( 1-e^{-t/\tau_\mathrm{r}} \right).
\end{equation}
The slow response contains low-frequency information for resonance frequency deviations. The speed of the response above is limited by $\tau_r$. For resonators with high quality factors, this response time can become very long. 
     
    One can alternatively extract the frequency information from the phase $\Delta \theta(t)$ via differentiation (with respect to time), as it has been shown in \cite{FF_diff1, FF_diff2}. This results in a {\it fast} but transient response as follows
    \begin{equation}
        \Delta \omega_\mathrm{fr}(t)= 
        \frac{d \Delta \theta(t)}{dt} =
        \Delta \omega_\mathrm{r} 
        e^{-t/\tau_\mathrm{r}}.
    \end{equation}
    The fast response contains high-frequency information for resonance frequency deviations but suppresses low-frequency phenomena such as thermal drift, due to differentiation.
    By combining (adding) the slow and fast responses, we obtain 
    \begin{equation}
        \Delta \omega (t)= 
        \frac{\Delta \theta(t)}{\tau_\mathrm{r}}+ 
        \frac{d \Delta \theta(t)}{dt}
        = \Delta \omega_\mathrm{r}
        \label{eq:ff_resp}
    \end{equation}
as an {\it instantaneous} and non-transient frequency response, which contains both low and high frequency information for resonance frequency deviations. This instantaneous response will only be smoothed and slowed down by any band-limiting mechanism, e.g., a low-pass filter, in the phase detector. For instance, when phase difference detection is performed with an I/Q demodulator as in a lock-in amplifier setup, the response speed will be determined by the low-pass filters in the demodulator represented by a transfer function $H_\mathrm{L}(s)$. The bandwidth for these filters needs to be smaller than (twice) the resonance frequency in order to filter out the high frequency (at twice the resonance frequency) signal components that are produced by the multipliers in the demodulator. By transforming equation (\ref{eq:ff_resp}) into the Laplace domain and including $H_\mathrm{L}(s)$ to represent the band-limited nature of phase detection, we obtain
\begin{equation}
    H_\mathrm{FF}(s) = \left(\frac{1}{\tau_\mathrm{r}}+ s \right)H_\mathrm{L}(s),
\end{equation}
which can be also written as
    \begin{equation}
    H_\mathrm{FF}(s)
       = \frac{1}{\tau_\mathrm{r}}\frac{1}{H_\mathrm{r}(s)}H_\mathrm{L}(s),
\end{equation}
where $H_\mathrm{r}(s)$ is a single-pole low-pass filter with the time
constant of the resonator
\begin{equation}
  H_\mathrm{r}(s)=\frac{1}{1+s \tau_\mathrm{r}},
\end{equation}
capturing the input-output frequency-domain response of the resonator. $H_\mathrm{FF}(s)$ can be used in computing the frequency-domain frequency response of the FF scheme (with combined fast and slow response) to step changes in the resonance frequency, as well as for characterizing frequency fluctuations due to noise.  

The frequency fluctuations caused by noise in the feedback free scheme are directly determined by the phase fluctuations of the input signal to the phase detector. The phase fluctuations detected by the phase detector can be expressed in the Laplace domain as
\begin{equation}
    \Delta \theta(s)= \theta_\mathrm{th}(s)H_\mathrm{r}(s)+\theta_\mathrm{d}(s).
\end{equation}
The thermomechanical phase fluctuations are shaped by the resonator characteristics before they are detected, while detection noise is fed into the phase detector unaltered \cite{Demir}. The frequency fluctuations of the FF tracking scheme are then obtained by simply multiplying the phase fluctuations by the transfer function of the FF scheme
\begin{equation}
    \Delta \omega (s) = \Delta \theta(s) H_\mathrm{FF}, 
\end{equation}
yielding the transfer functions for the two noise sources:
    \begin{equation}
    \begin{split}
        H_\mathrm{\theta_\mathrm{th}}^\mathrm{FF}(s) = \frac{1}{\tau_\mathrm{r}}H_\mathrm{L}(s),\\
        H_\mathrm{\theta_\mathrm{d}}^\mathrm{FF}(s) = \frac{1}{\tau_\mathrm{r}}\frac{1}{H_\mathrm{r}(s)} H_\mathrm{L}(s).
    \end{split}
    \end{equation}

\subsection{Self-sustaining oscillator (SSO)}
    
    \begin{figure}
        \centering
        \includegraphics[clip,width=\columnwidth]{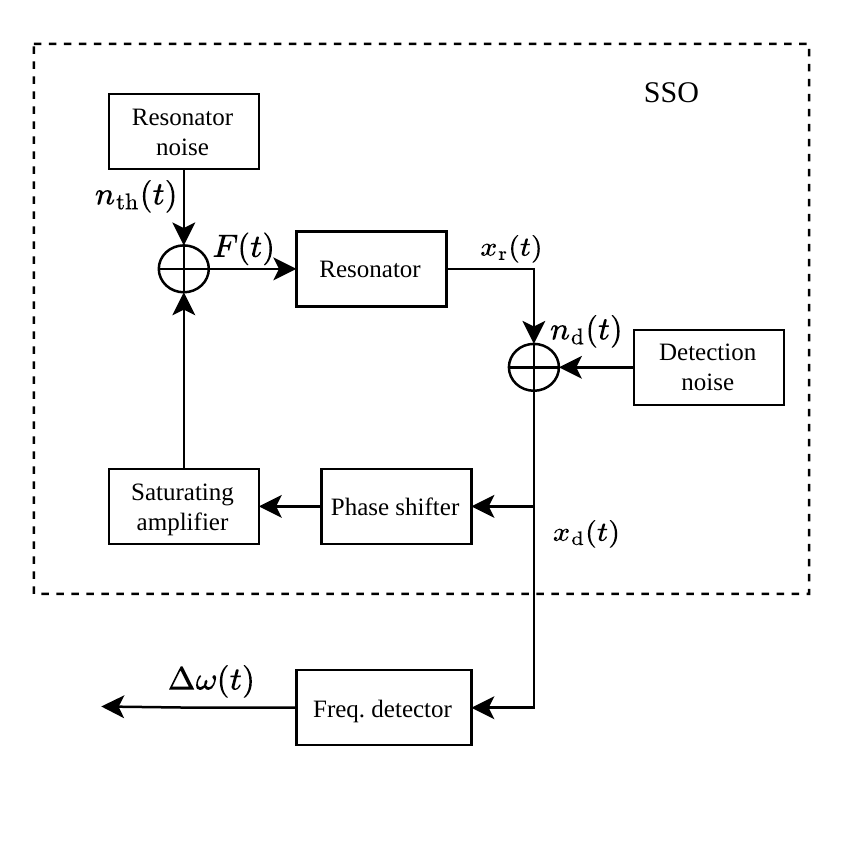}
        \caption{A block diagram showing a standard SSO scheme, with a phase shifter and a non-linear saturating element in the positive feedback path. The thermomechanical noise force $n_\mathrm{th}(t)$ is added at the input of the resonator. The detection noise $n_\mathrm{d}(t)$ is added in the transduction of the displacement signal $x_\mathrm{r}(t)$ into an electrical signal $x_\mathrm{d}(t)$. The electrical signal is then filtered and acquired by the frequency detector.}
        \label{fig:sso_noise}
    \end{figure}
    
    The standard SSO configuration as shown in Figure~\ref{fig:sso_noise} is based on the Barkhausen criterion, which requires that the closed-loop gain is equal to one. This can be achieved by introducing a non-linearity (saturating amplifier) $h( \cdot )$ in the loop  which stabilizes the amplitude. This non-linearity also generates higher-order harmonics, which are well filtered if the resonator has a large quality factor. The Barkhausen criterion also imposes a phase condition for stable and sustained oscillations: The phase around the loop needs to be $n 2\pi$ where $n = 1, 2...$ is much smaller than the quality factor of the resonator. The phase condition is realized by introducing a phase shifting element that phase shifts or delays the signal at the output of the resonator to generate the feedback drive. 
    
    The working principle of a standard SSO is described in \cite{Demir}, where a model was derived.  We start with the following 
    equation from \cite{Demir}  
    \begin{equation}
    \begin{split}
        \frac{d}{dt}\Delta \theta_\mathrm{r}(t) 
        & =\frac{1}{\tau_\mathrm{r}} \frac{1}{A_\mathrm{rss}} \frac{Q}{m\omega_\mathrm{r}^2} [h(A_\mathrm{rss})n_\mathrm{d}(t) + n_\mathrm{th}(t)] \\ 
        & = \frac{1}{\tau_\mathrm{r}} \left[ \frac{Q}{m\omega_\mathrm{r}^2} h(A_\mathrm{rss})\theta_\mathrm{d}(t) + \theta_\mathrm{th}(t) \right],
    \end{split}
    \label{phasenoisesso_x}
    \end{equation}
    with the steady-state and noiseless amplitude $A_{rss}$. The equation above describes the fluctuations caused by noise at the resonator output phase, hence in the signal $x_\mathrm{r}(t)$ in Figure \ref{fig:sso_noise}. Due to the feedback path, detection noise $n_\mathrm{d}(t)$ contributes to the phase fluctuations of $x_\mathrm{r}(t)$. However, phase fluctuations in the detected resonator output ($x_\mathrm{d}(t)$ in Figure \ref{fig:sso_noise}) have an additional contribution due to detection noise. Equation (\ref{phasenoisesso_x}) can be augmented as follows to derive the equation for phase fluctuations in $x_\mathrm{d}(t)$
    \begin{equation}
        \frac{d}{dt}\Delta \theta(t) 
        = \frac{1}{\tau_\mathrm{r}} \left[ \frac{Q}{m\omega_\mathrm{r}^2} h(A_\mathrm{rss})\theta_\mathrm{d}(t) + \theta_\mathrm{th}(t) \right]+ \frac{d}{dt} \theta_\mathrm{d}(t).
        \label{phasenoisesso_y}
    \end{equation}
    As stated in \cite{Demir}, the gain condition for a self sustaining oscillator  
    \begin{equation}
        \frac{Q}{m\omega_\mathrm{r}^2} h(A_\mathrm{rss})=1
    \end{equation}
    needs to be met, which yields 
    \begin{equation}
        \frac{d}{dt}\Delta \theta(t) 
        = \frac{1}{\tau_\mathrm{r}} \left[ \theta_\mathrm{d}(t) + \theta_\mathrm{th}(t) \right]+ \frac{d}{dt} \theta_\mathrm{d}(t).
    \end{equation}
    The transfer functions from thermomechanical and detection noise sources to the (frequency of the) output $x_\mathrm{d}(t)$ can be derived based on the equation above
    \begin{equation}
    \begin{split}
        H_\mathrm{\theta_\mathrm{th}}^\mathrm{SSO}(s) = \frac{1}{\tau_\mathrm{r}}H_\mathrm{L}(s),\\
        H_\mathrm{\theta_\mathrm{d}}^\mathrm{SSO}(s) = \frac{1}{\tau_\mathrm{r}}\frac{1}{H_\mathrm{r}(s)} H_\mathrm{L}(s),
    \end{split}
    \end{equation}
    where $H_\mathrm{r}$ is a single-pole low-pass filter with the time constant of the resonator. $H_\mathrm{L}$ 
    has low-pass characteristics and represents the bandwidth limiting (noise filtering) mechanisms in the frequency detection device.
    
\subsection{Phase-locked loop oscillator (PLLO)}
    \begin{figure}
        \centering
        \includegraphics[width=\columnwidth]{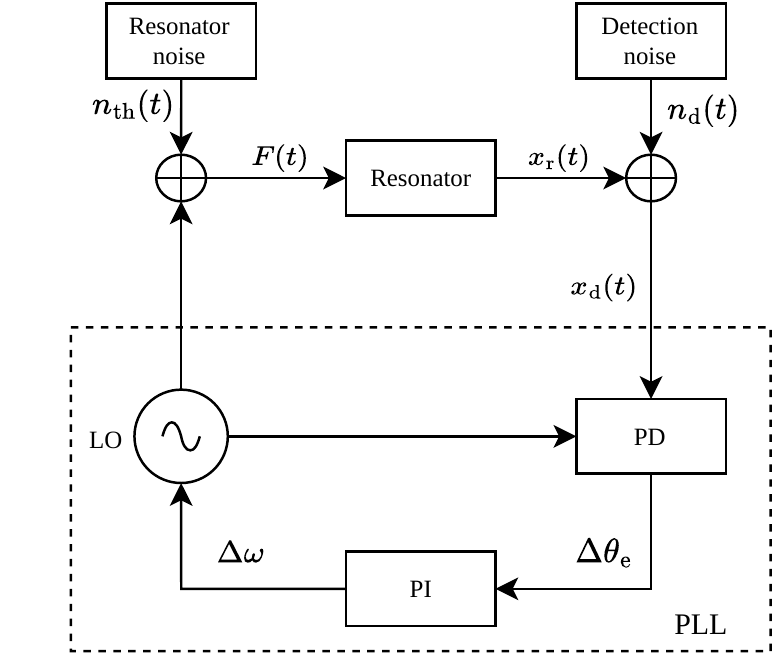}
        \caption{The phase detector (PD) computes the phase difference between the resonator response and the local oscillator (LO), which (minus a set point) is fed into the PI controller as an error signal. PI controller tunes the frequency of the LO, which is used to drive the resonator. }
        \label{fig:pll_block}
    \end{figure}
    The phase-locked loop oscillator (PLLO) essentially is a closed-loop extension of the FF scheme, which continually updates the drive frequency to match the resonators' resonance frequency. The block diagram in Figure \ref{fig:pll_block} shows the standard PLLO setup with a simplified representation of its internal structure. PLLO is usually realized digitally, for instance by using a lock-in amplifier with an integrated PLL. The internal oscillator, also called a controlled or local oscillator (LO), of the PLLO drives the resonator. The thermomechanical noise $n_\mathrm{th}$ is added to the drive at the input of the resonator. The motion of the resonator $x_\mathrm{r}(t)$ is transduced into an electrical signal $y(t)$. The electrical signal then goes into a phase detector (PD) and is compared to the internal LO signal. PD is realized with an I/Q demodulator, and has internal low-pass filters (LPF) with a cut-off frequency $f_\mathrm{L}$ (and time constant $\tau_\mathrm{L}=\tfrac{1}{2\pi f_\mathrm{L}})$. This filter removes high frequency signal components, as well as high frequency noise, but limits the bandwidth of phase detection. The output of PD (minus a set point) is used as the error signal $\Delta \theta_\mathrm{e}$ that represents the difference in phase between the resonator response and the LO drive signal. The PI controller produces the control signal that sets the frequency of the LO. The negative feedback loop maintains the desired phase difference between the resonator response and the LO. The speed with which the PI controller regulates the oscillator is characterized in \cite{silvanbook} using the system bandwidth of the PLLO or $f_\mathrm{PLL}$ \cite{Pedram}. The PI coefficients are calculated from the desired system bandwidth  as follows \cite{Demir}
    \begin{equation}
    \begin{gathered}
        k_\mathrm{p} = 2\pi f_\mathrm{PLL} = \frac{1}{\tau_\mathrm{PLL}}, \\
        k_\mathrm{i}= \frac{k_p}{\tau_\mathrm{r}},
    \end{gathered}
    \label{eq:pid_coef}
    \end{equation}
    where $k_\mathrm{p}$ is the proportional, and $k_\mathrm{i}$ is the integral coefficient of the PI controller. The noise transfer functions for the PLLO can be derived based on the loop dynamics as described in \cite{Demir}:
    \begin{equation}
    \begin{gathered}
        H_\mathrm{\theta_\mathrm{th}}^\mathrm{PLL}(s) = \frac{1}{\tau_\mathrm{r}} 
        \frac{(sk_\mathrm{p}+k_\mathrm{i})H_\mathrm{L}(s)}{s^2+\frac{s}{\tau_\mathrm{r}}+(sk_\mathrm{p}+k_\mathrm{i})H_\mathrm{L}(s)},\\
        H_\mathrm{\theta_\mathrm{d}}^\mathrm{PLL}(s) = \frac{1}{\tau_\mathrm{r}}\frac{1}{H_\mathrm{r}(s)} 
        \frac{(sk_\mathrm{p}+k_\mathrm{i})H_\mathrm{L}(s)}{s^2+\frac{s}{\tau_\mathrm{r}}+(sk_\mathrm{p}+k_\mathrm{i})H_\mathrm{L}(s)}.
    \end{gathered}
    \end{equation}
    With the parameters in (\ref{eq:pid_coef}), the noise transfer functions above take the following simpler forms \cite{Demir}
\begin{equation}
    \begin{gathered}
        H_\mathrm{\theta_\mathrm{th}}^\mathrm{PLL}(s) = \frac{1}{\tau_\mathrm{r}} 
        \frac{H_\mathrm{L}(s)}{H_\mathrm{L}(s) + s\,\tau_\mathrm{PLL}},\\
        H_\mathrm{\theta_\mathrm{d}}^\mathrm{PLL}(s) = \frac{1}{\tau_\mathrm{r}}\frac{1}{H_\mathrm{r}(s)} 
        \frac{H_\mathrm{L}(s)}{H_\mathrm{L}(s) + s\,\tau_\mathrm{PLL}}.
    \end{gathered}
    \end{equation}

 \subsection{Allan deviation}
    The standard and well established method for characterizing frequency fluctuations is the Allan deviation $\sigma_y(\tau)$ \cite{silvanbook,Demir, Allan_Paper}. It is the square root of Allan variance that can be computed with 
    \begin{equation}
        \sigma_y^2(\tau) =\frac{1}{2\,(N-1)} \sum_{i=1}^N (y_{i+1,\tau}-y_{i,\tau})^2,
        \label{eq:avar1}%
    \end{equation}
    where $y_i$ is the $i^{\mathrm{th}}$ sample of averaged frequency over averaging time $\tau$, i.e.,
    \begin{equation}
        y_{i,\tau} = \frac{1}{\tau}\int_{(i-1)\tau}^{i \tau}y(t)\mathrm{d}t.
        \label{eq:avar2}%
    \end{equation}
    The frequency values when computing an Allan deviation need to be normalized, resulting in a fractional frequency 
    \begin{equation}
        y(t)=\frac{\Delta\omega(t)}{\omega_0}
    \end{equation}
Allan variance can also be computed in the frequency domain if the power spectral density of fractional frequency fluctuations is known:
    \begin{equation}
        \sigma_\mathrm{y}^2(\tau)= \frac{1}{2\pi}\frac{8}{\tau^2} \int_0^\infty \frac{[\sin(\frac{\omega\tau}{2})]^4}{\omega^2} S_\mathrm{y}(\omega)\mathrm{d}\omega.
        \label{eq:sigma}
    \end{equation}
    For white frequency fluctuations with $S_y(\omega)=\text{constant}$, (\ref{eq:sigma}) reduces to
    \begin{equation}\label{eq:one-over-sqrttau}
        \sigma_y^2=\frac{S_y(0)}{2 \tau}.
    \end{equation}
    Hence, a system limited, e.g., by thermal white noise, the resulting Allan deviation exhibits a $\sigma_y\propto 1/\sqrt{\tau}$ dependence with averaging time $\tau$. 
    
The power spectral density of frequency noise $S_{\Delta \omega}(\omega)$ can be computed as a superposition of the power spectral densities of thermomechanical and detection phase noise (\ref{eq:phasenoise}) multiplied with their corresponding transfer functions (magnitude squared) derived above \cite{Demir}, which then readily yields the fractional frequency noise required to compute the Allan deviation
\begin{equation}
\begin{split}
    S_\mathrm{y}(\omega)&= \frac{S_{\Delta \omega}(\omega)}{\omega_0^2}\\
    &= \frac{S_{\theta_\mathrm{th}}(\omega)|H_{\theta_\mathrm{th}}(\mathrm{j}\omega)|^2+ S_{\theta_\mathrm{d}}(\omega)|H_{\theta_\mathrm{d}}(\mathrm{j}\omega)|^2}{\omega_0^2}\\
    &= \frac{S_{\theta_\mathrm{th}}(\omega)}{\omega_0^2}\left[|H_{\theta_\mathrm{th}}(\mathrm{j}\omega)|^2 + \mathcal{K}^2|H_{\theta_\mathrm{d}}(\mathrm{j}\omega)|^2 \right].
\end{split}
\label{eq:S(omega)}
\end{equation}
    
\section{Methods}
The models for the three frequency tracking schemes were tested experimentally with a nanoelectromechanical system (NEMS) resonator. While the FF and PLLO tracking schemes were implemented with a commercial lock-in amplifier (HF2LI 50 MHz Lock-in Amplifier from Zurich Instruments), the SSO scheme was realized  with a frequency tracking prototype device from Invisible-Light Labs GmbH.

\subsection{NEMS resonator}\label{mag_setup}
\begin{figure}
        \centering
        \includegraphics[width=0.7\columnwidth]{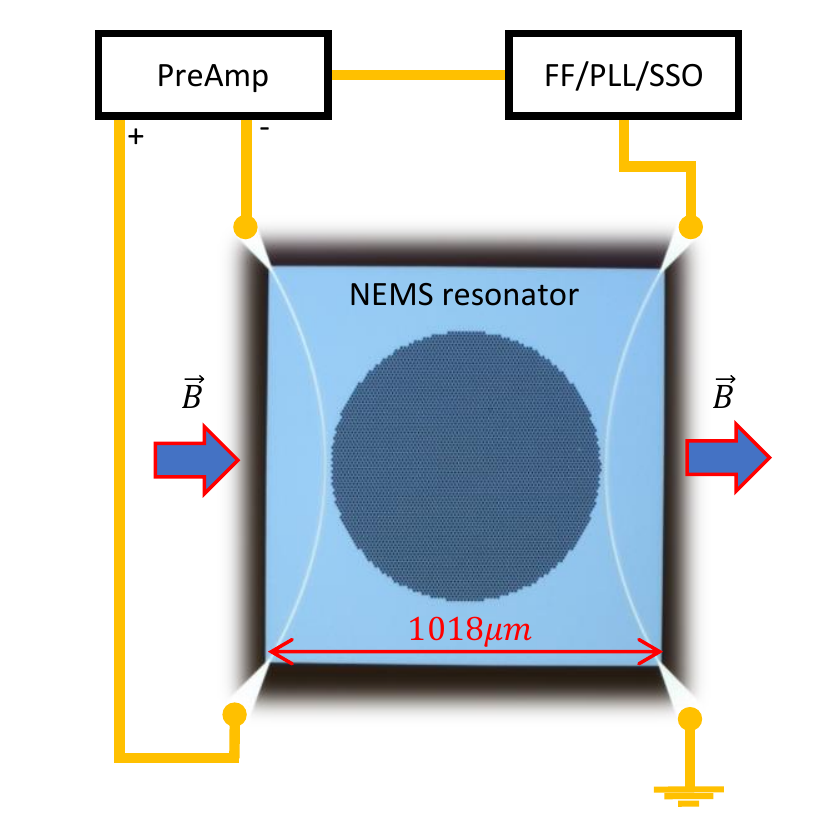}
        \caption{Schematic of the NEMS resonator used for all experiments featuring electrodynamic transduction. The circular area in the center where the SiN drumhead is perforated has no relevance for this work.}
        \label{fig:Chip}
    \end{figure}

The NEMS resonator used in this work (see Figure~\ref{fig:Chip}) consists of a \SI{1018}{\micro\meter} sized square membrane shaped resonator made of \SI{50}{\nano\meter} thick low-stress silicon-rich silicon nitride (fabricated by low pressure chemical vapour deposition). 
The electrical transduction is realized by two \SI{5}{\micro\meter} wide Au traces passing on the resonator. The membrane is placed in the center of a static magnetic field of about \SI{0.8}{\tesla}, created by a Halbach array of neodymium magnets, with the traces oriented perpendicular to the magnetic field. Exploiting the resulting Lorentz force, the metal traces can be used both to drive with an AC current, and in return, to detect the motion of the resonator through the magnetomotively induced voltage. The signal from the metal trace for detection is amplified with a low-noise differential pre-amplifier with a gain factor of $10^4$. The NEMS chip is placed in a vacuum chamber featuring a rotary vane pump, reaching a vacuum of $5.2\cdot10^{-3}\,\mathrm{mbar}$. With a resonance frequency of $\omega_r=82.3$~kHZ and a quality factor of $Q=97000$, the NEMS resonator has a response time (\ref{eq:responsetime}) of $\tau_\mathrm{r}= 2Q/\omega_r=0.4$~s.
    
    
\subsection{SSO with pulsed drive}
\begin{figure}
\centering
\includegraphics[width=0.8\columnwidth]{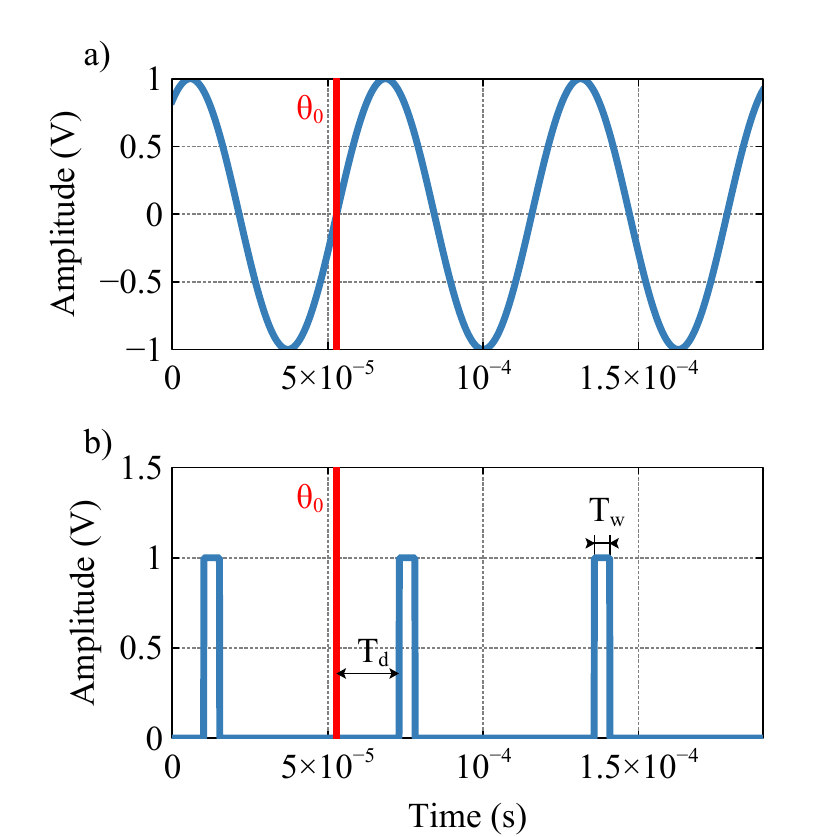}
\caption{Timing of the positive feedback system: (a) The sinusoidal signal at the resonator output is fed into the positive feedback mechanism. The mechanism detects the phase $\theta_\mathrm{0}$. (b) The pulse generation mechanism generates a pulse with a time delay $T_\mathrm{d}$ with respect to $\theta_\mathrm{0}$.}
\label{fig:timediag}
\end{figure}

The SSO frequency tracking system from Invisible-Light Labs that was used in this work is based on pulsed positive feedback \cite{dominguez2007influence,COLINET2004118}, producing pulses of width $T_{\mathrm{w}}$ with an adjustable delay $T_{\mathrm{d}}$ with respect to the signal phase $\theta_\mathrm{0}$, and a feedback-controlled amplitude to sustain a constant vibrational amplitude of the resonator. A schematic of the timing of the wave-forms at the outputs produced by the pulsed positive feedback is shown in Figure~\ref{fig:timediag}.

A narrow pulse with width $T_{\mathrm{w}}$ in the time domain corresponds to a sinc function in the frequency domain with zero points at integer multiples of $f_\mathrm{b} = \tfrac{1}{T_{\mathrm{w}}}$. The first zero point can be considered as the bandwidth of the pulse. The frequency content above $f_\mathrm{b}$ is small in comparison to the ones below $f_\mathrm{b}$ and do not contribute much to the signal. Hence, when actuating a resonator, a pulsed signal can be used instead of the typical sinusoidal signal, as it is common, e.g., for the characterization of tuning forks \cite{elektrischemesstechnik}. 

\begin{figure}
\centering
\includegraphics[width=\columnwidth]{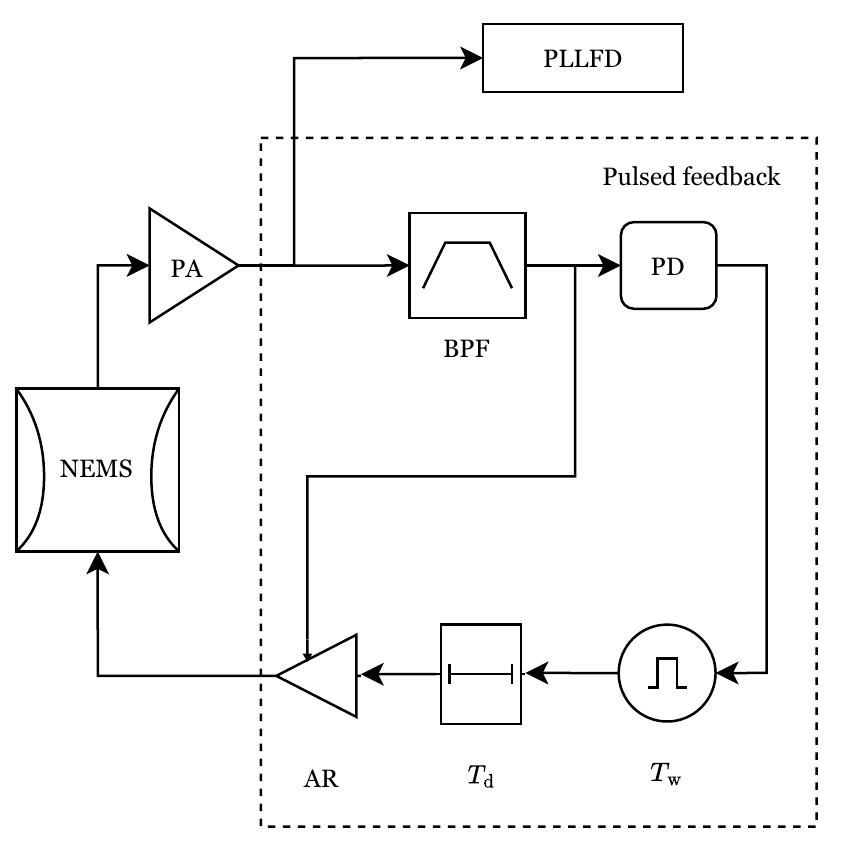}
\caption{Block diagram of the pulsed positive feedback system provided by Invisible-Light Labs used for the SSO scheme. The NEMS resonator response is amplified by a pre-amplifier (PA) and filtered by a band-pass filter (BPF). The phase is then detected by a phase detector (PD). The pulse generated by the pulse generator is delayed by the delay line. The pulses go through a feedback-controlled amplitude regulation (AR) to drive the NEMS resonator at a fixed vibrational amplitude.}
\label{fig:PosFedBlock}
\end{figure}

Figure~\ref{fig:PosFedBlock} shows the block diagram representation of the positive feedback system used in this work. The signal generated by the NEMS resonator is first amplified with a pre-amplifier before it enters the positive feedback path. The feedback signal is first passed through a band-pass filter, which serves two purposes. First, it reduces detection noise, and second, it attenuates  unwanted modes of the resonator. Afterwards, the signal passes through a phase detector, which is able to detect a phase at $0^\circ$ or $180^\circ$. Phase shifts introduced by components in the loop are compensated by an adjustable time delay element. It induces a delay $T_\mathrm{d}$ to the output pulse. $T_\mathrm{d}$ needs to be large enough to cover a phase delay between $0^\circ$ and $180^\circ$, which requires a maximum delay of $T_\mathrm{d, max} \geq \frac{\pi}{\omega_\mathrm{r} }$. The driving pulse is generated via a pulse generation mechanism that is triggered by the phase detector. The positive feedback system generates a pulse with adjustable width $T_\mathrm{w}$. Modification of the resonance frequency and therefore the pulse frequency at constant pulse width will result in a change in the energy pumped into the system and hence the amplitude of the resonator will change. This behavior is compensated by regulating the amplitude of the pulse. The amplitude regulation adjusts the pulse height in accordance with the measured input signal level and desired set-point.

The amplitude control is performed by the amplitude regulation block. It measures the amplitude of the signal at the output of the band-pass filter and compares it to the desired set-point. 

To adjust the output voltage to the required NEMS voltage values, a $100\,\mathrm{dB}$ attenuator is placed at the output of the positive feedback. The band-pass filter, amplitude regulation, $T_\mathrm{d}$, and $T_\mathrm{w}$ are adjustable parameters. 

\subsection{Tracking schemes setup}
For the PLLO scheme, the desired system target bandwidth is determined by the PI coefficients with the relations in equation (\ref{eq:pid_coef}). The low-pass filter bandwidth $f_\mathrm{L}$ of the PLLO demodulator is set five times larger than the target bandwidth. The sampling rate is set at least ten times larger than the LPF bandwidth. 
    
In order to compare the tracking schemes, the sensor needs to be driven at its resonance. In the case of the SSO, the delay of the pulse was adjusted until the phase of the pulses match the phase of the resonator. This is obtained when the maximum oscillation amplitude is reached for a fixed pulse amplitude.  In the case of the PLLO, to lock onto the phase that corresponds to the resonance frequency, a frequency sweep needs to be performed. This sweep is also needed to determine the resonance, and hence the drive frequency for the FF scheme. To make sure that the resonator has the same vibrational amplitude for all tracking schemes, the drive amplitude was set so that in all schemes the output of the pre-amplifier had the same amplitude of \SI{22.1}{\milli\volt}. The PLLO dynamics is determined by the chosen target bandwidth. The dynamics of the FF approach is limited by the demodulation filter bandwidth in the phase detector. While the SSO core itself exhibits an instantaneous response to sudden resonance frequency changes, the dynamics are limited by the frequency detection device that is used in conjunction with the SSO \cite{Demir}. 

While the pulsed positive feedback device comes with a built-in frequency counter, in this work to provide a one to one comparison between the different schemes, a PLL-based frequency detector (PLLFD) was used also for the SSO scheme. The controlling signal for the PLL's local oscillator is then used to measure the frequency of oscillation of the SSO core. This is in contrast with the use of a PLL in the PLLO-FLL scheme, where the local oscillator signal in fact drives the resonator and its frequency is locked to the resonance frequency by maintaining the appropriate phase difference between the resonator output and the drive. Using the same PLL frequency detection allows setting the system response time to be equal for all of the three tracking schemes.

\section{Results and discussion} \label{sec:measurements}

\subsection{Analytical results}
\begin{figure*}
\centering
\includegraphics[width=0.85\textwidth]{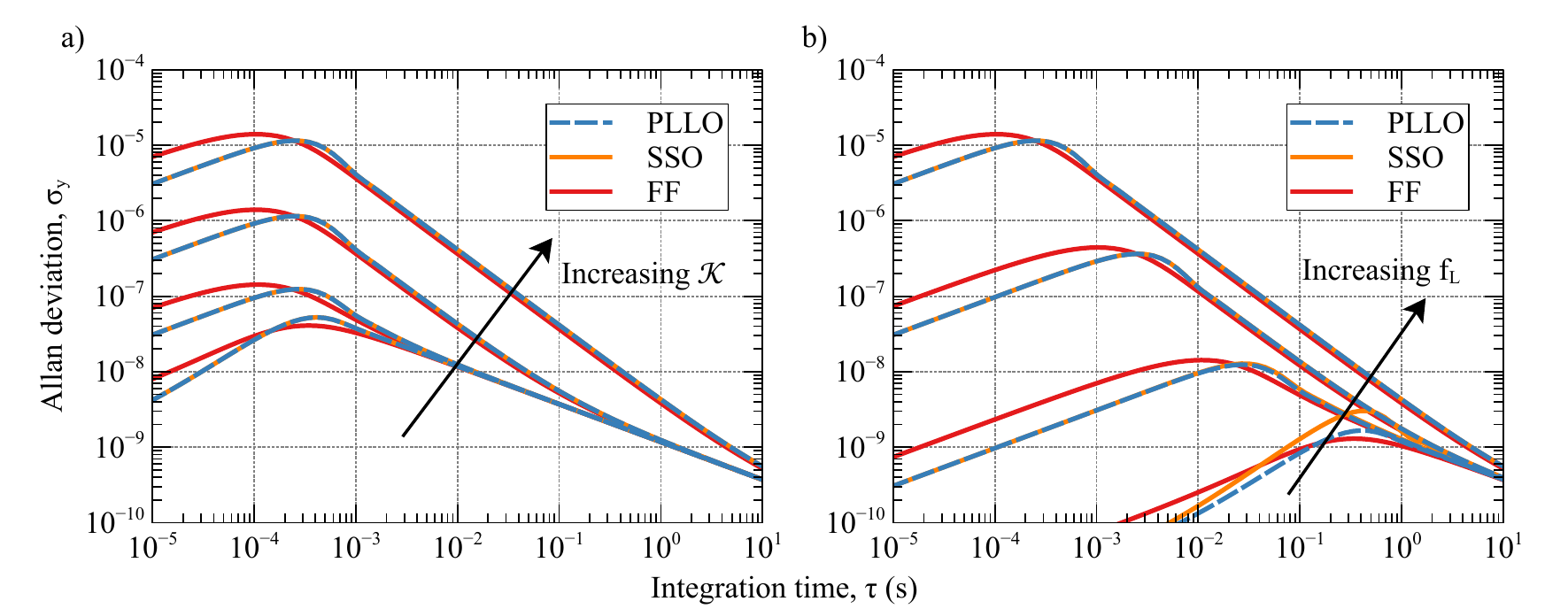}
\caption{Theoretical Allan deviations for the three frequency tracking schemes for (a) increasing detection noise ($\mathcal{K}= \{ 10^{-4}, 10^{-3}, 10^{-2}, 10^{-1} \}$) for a fixed bandwidth $f_\mathrm{L}=1$~kHz and (b) increasing system bandwidth ($f_\mathrm{L}= \{ 1\,\mathrm{Hz}, 10 \,\mathrm{Hz}, 100\,\mathrm{Hz}, 1\,\mathrm{kHz} \}$) for a fixed $\mathcal{K}=0.1$. All calculations were performed for a resonator time constant $\tau_\mathrm{r}=0.4$~s.}
\label{fig:allan_dev} 
\end{figure*}

Figure~\ref{fig:allan_dev} shows the theoretical Allan deviations for the three frequency tracking schemes.
Allan deviation for a NEMS resonator that is limited by thermomechanical and detection noise, exhibits two regimes with slopes proportional to $1/\tau$ and $1/\sqrt{\tau}$. A $1/\tau$ slope arises from (amplified) detection noise, whereas the more fundamental thermomechanical noise results in a $1/\sqrt{\tau}$ slope, as expected for a white noise source (\ref{eq:one-over-sqrttau}). 
(\ref{eq:S(omega)}) indicates that the functional form of the Allan deviation is determined by two factors. The first is the ratio between detection and thermomechanical noise $\mathcal{K}$.
As shown in Figure \ref{fig:allan_dev}a, increasing $\mathcal{K}$  increases the relative impact of detection noise, resulting in the change of the slope from $1/\sqrt{\tau}$ to the $1/\tau$. The second factor that affects the Allan deviation is the system bandwidth $f_\mathrm{L}$, which acts as a low-pass filter for signal variations. In all tracking schemes, which all use an I/Q demodulator for phase detection, signal variations are filtered by the low-pass filter inside the demodulator. In the PLLO scheme and in the PLLFD used in conjunction with the SSO, there is additional filtering due to the negative feedback loop dynamics with a bandwidth $f_\mathrm{PLL}$. In practice, the demodulator bandwidth is set larger than the loop bandwidth  ($f_\mathrm{L} \geq 5\,f_\mathrm{PLL}$). For the calculations presented in Figure~\ref{fig:allan_dev}, the bandwidths were chosen equal $f_\mathrm{L}=f_\mathrm{PLL}$ to obtain an equal overall bandwidth for all schemes. Since a first-order demodulator filter in PLLO and SSO-PLLFD effectively results in an overall second-order filter due to loop dynamics, a second-order demodulator filter is used in the FF scheme to obtain  the same effective filter order for all three schemes. 

Figure~\ref{fig:allan_dev}b shows the effect of changing the filter bandwidth $f_\mathrm{L}$ on the Allan deviation.  Decreasing $f_\mathrm{L}$ improves the filtering of detection noise and results in a smaller Allan deviation, however at the expense of a larger response time \cite{Demir,Pedram}. 

It is important to point out that even though thermomechanical noise is resolved above the detection noise background ($\mathcal{K}<1$) for all calculations presented in Figure~\ref{fig:allan_dev}, detection noise can nonetheless affect the resulting Allan deviations. As it is shown in Figure~\ref{fig:allan_dev}a, frequency fluctuations start to reach the  thermomechanical noise limit with a $1/\sqrt{\tau}$ slope not until $\mathcal{K}\leq 0.001$. This behavior can be ascribed to the long response time of the high-Q NEMS resonator. This finding underlines the importance of a low-noise readout that provides a highly resolved thermomechanical noise peak to obtain minimal frequency fluctuations with high-Q resonators .


\subsection{Experimental testing}
\begin{figure*}
    \centering
    \includegraphics[]{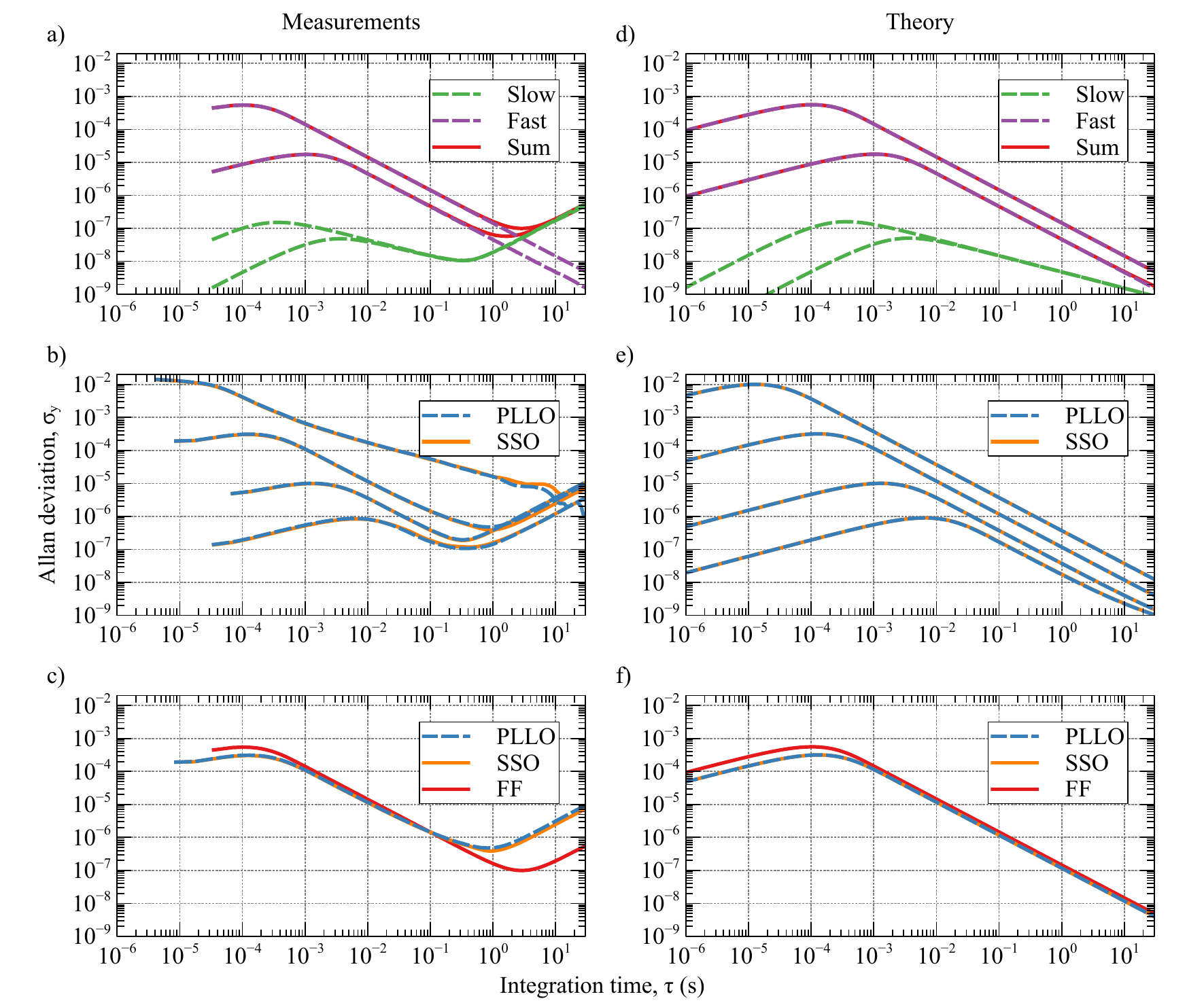}
    \caption{Comparison of experimental Allan deviations with theory. (a) Allan deviations for three FF tracking modes with second order demodulation filters of bandwidth $f_\mathrm{L}=\{100\,\mathrm{Hz}, 1\,\mathrm{kHz}\}$ (b) Comparison of PLLO and SSO tracking schemes for $f_\mathrm{PLL} = \{ 10\,\mathrm{Hz}, 50\,\mathrm{Hz}, 500\,\mathrm{Hz}, 5\,\mathrm{kHz} \}$, where a PLLFD is used as the frequency detector for the SSO scheme. c) Comparison of PLLO and SSO with $f_\mathrm{PLL}=500\,\mathrm{Hz}$ and FF scheme (combined/sum mode) with $f_\mathrm{L}=1\,\mathrm{kHz}$ d)-f) Theoretical model based computations using equations (\ref{eq:sigma}) and (\ref{eq:S(omega)}) for the same settings as in the measurements a)-c), and with calibration for detection noise level.}
    \label{fig:allan_dev_pll_vs_sso} 
    \end{figure*}

The frequency fluctuations for the different tracking schemes were studied by collecting the steady-state frequency data over one minute. The corresponding Allan deviation curves calculated from the frequency data are presented in Figures~\ref{fig:allan_dev_pll_vs_sso}a-c. The integrated electronic transduction of the NEMS resonator could not resolve its thermomechanical noise peak, resulting in $\mathcal{K}\gg 1$. The corresponding theoretical Allan deviation curves are plotted in Figures~\ref{fig:allan_dev_pll_vs_sso}d-f.

The FF measurements shown in Figures~\ref{fig:allan_dev_pll_vs_sso}a were performed with a second-order demodulation filter with the cut-off frequencies $f_\mathrm{L}=\{100\,\mathrm{Hz}, 1\,\mathrm{kHz}\}$. The experimental results show that the FF scheme in slow mode has a slope proportional to $1/\sqrt{\tau}$. Changing the filter bandwidth does not affect the precision performance of this scheme, because, in the slow mode, the limiting and determining factor is the mechanical time constant of the resonator. On the other hand, in the fast mode, the Allan deviation has a $1/\tau$ slope, pointing to the fact that precision is determined by (amplified) detection noise. With a larger filter bandwidth, the response becomes faster but at the expense of less precision indicated by a larger Allan deviation.  It can also be observed that the fast response, due to its high-pass but {\it low-stop} nature, does not contain information about slow processes like thermal drift. Combining the slow and fast response, according to (\ref{eq:ff_resp}), results in an Allan deviation that exhibits low-frequency phenomena while the response speed is limited only by the demodulation filter. The experimental Allan deviations can be recreated with high accuracy by the theoretical model, which is plotted in Figure~\ref{fig:allan_dev_pll_vs_sso}d.

For the PLLO and SSO-PLLFD schemes shown in Figure~\ref{fig:allan_dev_pll_vs_sso}b, measurements were performed for four different loop bandwidths $f_\mathrm{PLL}=\{ 10\,\mathrm{Hz}, 50\,\mathrm{Hz},  500\,\mathrm{Hz}, 5\,\mathrm{kHz} \}$, with a demodulation low-pass filter cut-off $f_\mathrm{L}=5\,f_\mathrm{PLL}$. The PI controller parameters were chosen according to (\ref{eq:pid_coef}). In this case, the closed-loop dynamics for the PLLFD of the SSO and the PLLO are almost equivalent, especially when $f_\mathrm{PLL}\,\tau_\mathrm{r} \gg 1$. Having the same filtering characteristics and response times for both tracking schemes, a fair comparison can be made. We observe that the Allan deviations for SSO-PLLFD and PLLO are almost identical for all system bandwidths. As the theory indicates, with increasing bandwidth, the amplification of detection noise becomes more severe and Allan deviations with $1/\tau$ dependence in the detection noise limited regime become worse. For $f_\mathrm{PLL}=5\,\mathrm{kHz}$ ($f_\mathrm{L}=25\,\mathrm{kHz}$) and a sampling rate of $230\,\mathrm{kSa/s}$, a premature $1/\sqrt{\tau}$ dependence can be observed with an even worse Allan deviation. This behavior is a result of the practical limitation of the digital PLL. This arises if the sampling rate is not at least 10 to 20 times larger than the demodulator low-pass filter bandwidth to prevent aliasing. It should be noted that, in the model based theoretical computations, it is assumed that the sampling rate is large enough to prevent aliasing in Allan deviation computations. There is almost perfect correspondence between measurements and model based computations, presented in Figure~\ref{fig:allan_dev_pll_vs_sso}e, for smaller bandwidths, when there is no aliasing. 
    
Figure~\ref{fig:allan_dev_pll_vs_sso}c shows the Allan deviations for the FF (combined/sum mode), PLLO, and SSO-PLLFD tracking schemes for a similar system bandwidth. Clearly all measured Allan deviation curves are practically identical. The theoretical model, shown in Figure~\ref{fig:allan_dev_pll_vs_sso}f, confirms that there is no difference in terms of performance between the three tracking schemes.
    

All experimental Allan deviations (except for fast mode FF) exhibit a thermal drift-related rise for large $\tau$. This is not present in the theoretical computations since thermal drift is not modeled.

\section{Conclusions} \label{sec:conclusions}
In this work, we extended the existing models for the FF and SSO tracking schemes. We showed that when the FF tracking scheme is operated in the combined/sum mode proposed in this work, it offers an equivalent speed versus accuracy trade-off characteristics as the closed-loop SSO and PLLO  schemes. This is achieved by combining (adding) the slow and fast frequency responses that can be obtained by simple processing of the phase output from the demodulator. We also demonstrated that the SSO scheme has the same frequency fluctuation performance as the PLLO scheme. Finally, we compared the FF, SSO, and PLLO tracking schemes and showed that all tracking schemes have equivalent steady-state frequency fluctuation performance. There are two main parameters that affect the performance of all tracking schemes: (1) the ratio between the thermomechanical noise peak and detection noise floor, $\mathcal{K}$, (2) the filtering properties of the detection device. We further showed that a self sustaining oscillator tracking scheme with pulsed drive performs perfectly according to the theoretical model of sinusoidal positive feedback. 
These results give the user an option to choose the tracking scheme based on cost, robustness, ease of implementation, and usability in practice instead of fundamental differences in performance. 
    
\section*{Acknowledgements} \label{sec:acknowledgements}
We would like to thank Andreas Kainz and Franz Keplinger for constructive discussions that gave us motivation for designing the system described in the paper. This work received funding from the European Innovation Council under the European Union's Horizon Europe Transition Open program (Grant agreement: 101058711-NEMILIES).

\bibliography{bibliography.bib}

\end{document}